\documentclass[journal]{IEEEtran}
\usepackage{graphicx}
\usepackage{xcolor}
\usepackage[normalem]{ulem}
\usepackage{hyperref}

\usepackage{amssymb, amsmath}

\usepackage{amsmath, bm}

\ifCLASSINFOpdf
\else
\fi
\usepackage{amsmath}
\usepackage{array}
\usepackage{fixltx2e}

\usepackage{stfloats}
\begin{document}
%
\title{ A Novel Exact Analytical Expression for the Magnetic Field of a Solenoid }
%
%
%

\author{\IEEEauthorblockN{Mostafa Behtouei$^1$,
Luigi Faillace$^1$,
Bruno Spataro$^1$,
Alessandro Variola$^1$
 and Mauro Migliorati$^{2,3}$
}

\IEEEauthorblockA{$^1$INFN, Laboratori Nazionali di Frascati, P.O. Box 13, I-00044 Frascati, Italy}

\IEEEauthorblockA{$^2$Dipartimento di Scienze di Base e Applicate per l'Ingegneria (SBAI), Sapienza University of Rome, Rome, Italy}

\IEEEauthorblockA{$^3$INFN/Roma1, Istituto Nazionale di Fisica Nucleare, Piazzale Aldo Moro, 2, 00185, Rome, Italy }

}

\maketitle

\begin{abstract}
In this paper we present the analytical calculations to derive the magnetic field of a solenoid by solving exactly a fractional integral with the use of a novel method. Starting from the Biot-Savart law, we consider a coil of negligible thickness with a stationary electric current. We derive the expressions of the on and off-axes magnetic field components. The results have been compared to some simplified and known analytical formulae as well as to a commercial numerical code showing a good agreement.

\end{abstract}



%
\IEEEpeerreviewmaketitle

\section{Introduction}
%
%
%
%
\IEEEPARstart{T}{he} evaluation of the magnetic field in a solenoid is important in many fields of applied physics. For example, it is used in  particle accelerators  to counteract the effect of beam blow-up, due to space charge, in high brilliance photoinjectors \cite{ferrario2007direct,migliorati2009transport} as well as for high gradient accelerating structures \cite{stratakis2010effects,nezhevenko200134,behtouei2020initial}.  Moreover the use of magnetic correctors and steering devices are largely used in linear accelerators to align the trajectory of the charged beam \cite{brady2013encyclopedia}. 

The basic laws of current-carrying coils, representing a solenoid, have been established experimentally by Biot and Savart and then developed by Ampere \cite{pappas1983original, jackson1999classical}.  The Biot-Savart law is a consequence of Maxwell's equations. The procedure to determine the magnetic field is as follows:
 
Magnetic fields generated by steady current can be obtained from the curl of the vector potential, 

 \begin{equation}\label{1}
\bf B={\bf \nabla} \times A
\end{equation}

Assuming the Maxwell's equations, choosing the Coulomb gauge, ${\bf \nabla \cdot A}$=0 and implying the steady state condition, gives 

 \begin{equation}
\nabla^2 {\bf A}=\mu_0 {\bf J}
\end{equation}

This is Poisson's equation and the solution is 

 \begin{equation}
{\bf A}(\mathrm{p})=\frac{\mu_0}{4\pi} \int \frac{{\bf J}(\mathrm{p'})}{|{\bf p}-{\bf p'}|} d^3\mathrm{p'}
\end{equation}

where ${\bf p}$ is the observation point and ${\bf p'}$ is the integration variable. From Eq. ($\ref{1}$) we finally obtain the Biot-Savart law for a wire of finite thickness

 \begin{equation}
{\bf B}(\mathrm{p})=\frac{\mu_0}{4\pi} \int \frac{{\bf J}(\mathrm{p'}) \times ({\bf p}-{\bf p'})}{| {\bf p}-{\bf p'}|^3} d^3\mathrm{p'}
\end{equation}
 where $d^3\mathrm{p'}$ is the infinitesimal volume element. For a thin wire with an infinitesimal element, $d\mathrm{p'}$, this reduces to
 
  \begin{equation}\label{5}
{\bf B}(\mathrm{p})=\frac{\mu_0}{4\pi} \int \frac{{\bf I}  \times ({\bf p}-{\bf p'})}{|{\bf p}-{\bf p'}|^3}d\mathrm{p'}.
\end{equation}

 In the past some methods have been investigated in order to solve the above integrals to have analytic expressions of the magnetic fields produced by a thin wire, a 2D surface element or, in a general form, by a current density.  The authors of \cite{bassetti1996analytical}  proposed an analytical solution starting from the numerical approach developed by  Caspi \cite{caspi1991m,caspi1992approach,caspi1994use} and Leleux \cite{leleux1986complements}, and they described the magnetic potential near the axis of a multipole with the inclusion of the magnetic dipole. However, the problem for the magnetic field components far from the axis needed the evaluation of a large number of terms for the expansion coefficients. The authors of \cite{caciagli2018exact} have derived an analytical expression for the magnetic field of a cylinder of finite length with a uniform, transverse magnetization by using the assumptions that there are no free currents and that the magnetic field could be expressed as the gradient of a magnetostatic scalar potential. The final solution of the magnetic field components has been written in terms of the complete elliptic integrals of the first, second and third kind. 

The problem that all authors were facing \cite{bassetti1996analytical,caspi1994use,caciagli2018exact}, was that the final integral in cylindrical coordinates of the magnetic field components or the vector potential become a fractional integral of the order 3/2 and 1/2, respectively. In both cases we have to face a branch line. In this paper we propose a solution of the problem by cutting the branch line in order to have an analytic function inside the integral instead of multi-valued operation (obtained in the case of branch line). This approach can also be used for calculating the radial force and its derivatives caused by the centrifugal space charge. The author of \cite{bassetti1986analytical} has derived an equation for the derivatives of the radial force which is related to the centrifugal space charge using the elliptical integral.

In the next section we will obtain an equation for the magnetic field generated by a single coil of radius $R$ and we will calculate analytically the field amplitude on and off axis in a point at an arbitrary distance $r$ from the axis of the coil.

\section{Derivation of magnetic induction generated by a wire carrying current (circular current loop)}

\begin{figure}
 \begin{center}
\includegraphics[width=0.9\linewidth]{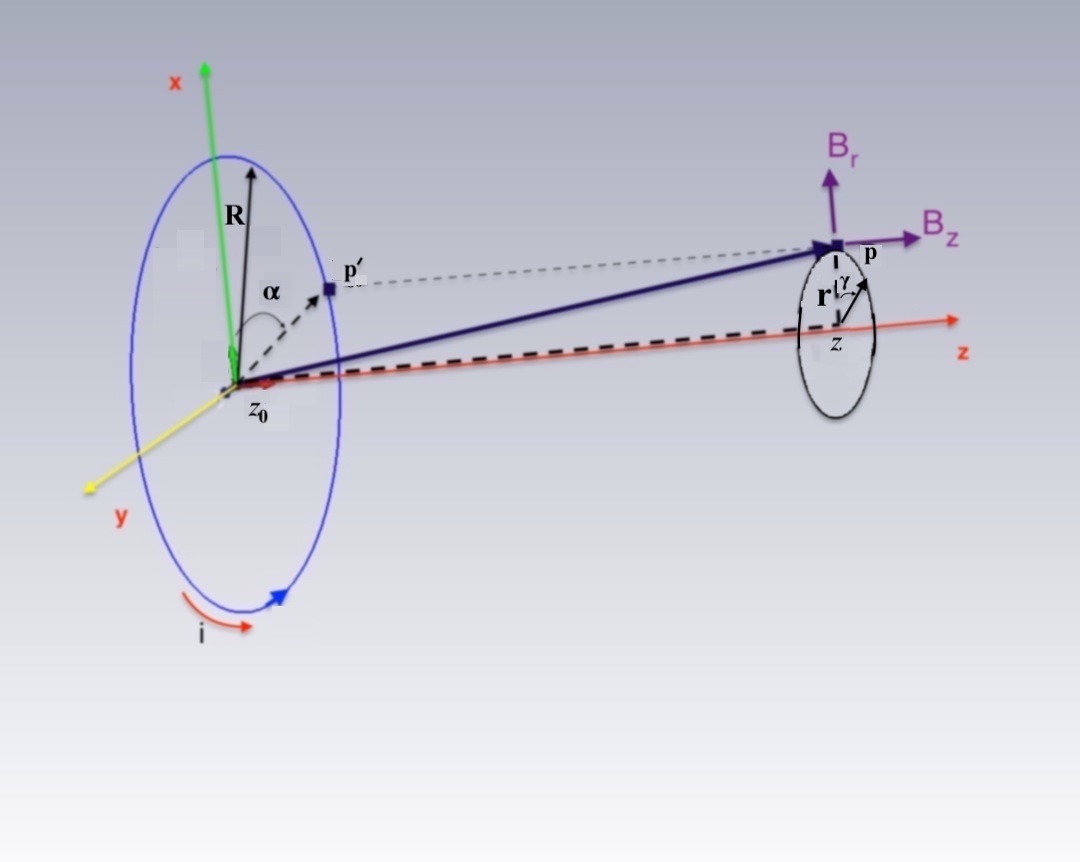}

\caption{ Off-Axis Field Due to a Current Loop}
\end{center}
{\small B is the magnetic field at any point in space out of the current loop.

$B_z$:  magnetic field component in the direction of  the coil axis. 

$B_r$:  radial magnetic field component.

$I$ is the current in the wire.

$R$ is the radius of the current loop.

z is the distance, on axis, from the center of the current loop to the field measurement point.

$r$ is the radial distance from the axis of the current loop to the field measurement point}

$\alpha$ denotes for the angle of the current element

$\gamma$ stands for the angle of the observer where the magnetic field components are to be calculated

      \end{figure}
      
Let ${\bf p'}$ and ${\bf p}$ be the positions of an infinitesimal element of a coil of radius $R$ in the position $z_0$ and of a point  located at a distance z from the coil and  $r$ from the z-axis, respectively. Their positions in cylindrical coordinates are Fig. 1:

\begin{equation}\label{7}
{\bf  p'}=(R\  \cos \alpha, R\  \sin \alpha, z_0)
\end{equation}

\begin{equation}\label{8}
{\bf  p}=(r\  \cos \gamma, r\  \sin \gamma, z)
\end{equation}

In the special case of a constant current $I$,  by calling $d \bm \sigma$ an infinitesimal element of coil pointing in the direction of current flow, Eq. ($\ref{5}$) can be written as

\begin{equation}\label{9}
{\bf  B} ( \mathrm{p})=\frac{\mu I}{4\pi} \int\ \frac{d {\bm \sigma} \times ({\bf  p}-{\bf  p'})}{|{\bf p}-{\bf p'}|^3}
\end{equation}
with,

\begin{equation}\label{10}
d  {\bm \sigma}=R\ d\alpha\  {\bm \tau}
\end{equation}
where ${\bm \tau}$ is the unit vector tangent to the circumference of radius $R$ and $d\alpha$ denotes the infinitesimal variation of the angle as shown in the figure. We can write ${\bm \tau}$ as:

\begin{equation}
{\bm \tau}=-\hat i \sin \alpha+\hat j \cos \alpha
\end{equation}
with $\hat{i}$ and $\hat{i}$ the unit vectors along $x$ and $y$. By replacing the above equation inside of the Eq. ($\ref{10}$) we obtain:

\begin{equation}\label{12}
d {\bm \sigma}=R\  d\alpha\ (-\hat i \sin \alpha+\hat j \cos \alpha)
\end{equation}

Subtracting the two vectors of the Eqs. ($\ref{7}$) and ($\ref{8}$) we have,

\begin{equation}\label{13}
{\bf p}-{\bf p'}=\hat i(r \cos \gamma-R \cos \alpha)+\hat j (r \sin \gamma-R \sin \alpha)+\hat k(z-z_0)
\end{equation}
and, replacing Eq. ($\ref{12}$)  into the Eq. ($\ref{9}$) we obtain:

\begin{equation}\label{14}
{\bf B} (\mathrm {p})=\frac{\mu_0  I R}{4 \pi} \int \frac{ {\bm \tau} \times  ({\bf p}-{\bf p'}) }{|{\bf p}-{\bf p'}|^3} \ d\alpha
\end{equation}
where cross product of  ${\bm \tau}$ and  $({\bf p}-{\bf p'})$ is:

  \begin{multline}\label{15}
  {\bm \tau} \times ({\bf p}-{\bf p'})=
\begin{vmatrix}
    \hat i&\hat j&\hat k \\
  -\sin \alpha&\cos \alpha& 0\\
  r\ \cos \gamma-R\ \cos \alpha & r\ \sin \gamma-R\ \sin \alpha&z-z_0
\end{vmatrix}\\
=\hat i\ (z-z_0) \ \cos \alpha + \hat j\ (z-z_0)\ \sin \alpha+ \hat k\ (R-r\ \cos(\gamma-\alpha))
\end{multline}

To solve the integral of  Eq. ($\ref{14}$) we can calculate  $|{\bf p}-{\bf p'}|^2$ as follows:

\begin{equation}\label{16}
 |{\bf p}-{\bf p'}|^2=r^2+R^2+ (z-z_0)^2-2rR \cos (\gamma-\alpha)
\end{equation}

Substituting Eqs.  ($\ref{15}$) and ($\ref{16}$) into the Eq. ($\ref{14}$) we obtain:

\begin{multline}
{\bf B} (\mathrm{p})=\frac{\mu_0 { I} R}{4 \pi} \\
\label{137}
\int  \frac{(\hat{i} \cos \alpha + \hat{j}\sin \alpha)(z-z_0)+ \vec{k}\ (R-r\ \cos(\gamma-\alpha))}{[r^2+R^2+ (z-z_0)^2-2rR \cos (\gamma-\alpha)]^{3/2}}\ d\alpha
\end{multline}

Writing the components of the magnetic field along the x, y and z coordinates gives:

\begin{multline}\label{18}
 \mathrm{B}_x=\frac{\mu_0 I R (z-z_0)}{4 \pi} \\
\int_0^{2\pi}  \frac{ \cos\alpha }{[r^2+R^2+ (z-z_0)^2-2rR \cos(\gamma-\alpha)]^{3/2}}\ d\alpha
\end{multline}

\begin{multline}\label{19}
\mathrm{B}_y=\frac{\mu_0  I R (z-z_0)}{4 \pi} \\
\int_0^{2\pi}   \frac{ \sin \alpha }{[r^2+R^2+ (z-z_0)^2-2rR \cos(\gamma-\alpha)]^{3/2}}\ d\alpha
\end{multline}

\begin{multline}
 \mathrm{B}_z=\frac{\mu_0 I R }{4 \pi} \\
\int_0^{2\pi}   \frac{ (R-r\cos(\gamma-\alpha)) }{[r^2+R^2+ (z-z_0)^2-2rR\ \cos(\gamma-\alpha)]^{3/2}}\ d\alpha    
\end{multline}

Let D be the boundary of $C^3$, that is,

\begin{equation}
\mathrm{D}=\{ {\bf p}=\{x, y, z\} \in C^3 |  \sqrt{x^2+y^2} < R \}
\end{equation}

We apply the above boundary because we are working on a cylindrical shape type of solenoid and the magnetic field outside of the solenoid is not considered.

The basic differential  laws of magnetostatics are given by:

\begin{equation}
\nabla \times {\bf B} ( {\bf p})=\frac{4\pi}{c} \ {\bf j}, \ \ {\bf p} \in D
\end{equation}

\begin{equation}\label{23}
\nabla \cdot {\bf B} ({\bf p})=0, \  \  {\bf p} \in D
\end{equation}

To verify Eq. ($\ref{23}$) we rewrite the magnetic field vector in cylindrical coordinates:

\begin{equation}\label{24}
 \mathrm{B}_r={\bf B}\ \cdot \ {\bf r}=  B_x\ \cos \gamma+  B_y\ \sin \gamma
\end{equation}

\begin{equation}\label{25}
 \mathrm{B}_{\gamma}={\bf B}\ \cdot \ {\bm \tau}=- B_x\ \sin\gamma+  B_y\ \cos \gamma
\end{equation}
where,

\begin{equation}
{\bm \tau}=(-\sin \gamma,\  \cos \gamma,\ 0)
\end{equation}

\begin{equation}
{\bf r}=(\cos \gamma,\  \sin \gamma,\ 0)
\end{equation}

By substituting Eqs. ($\ref{18}$), ($\ref{19}$) into the Eqs. ($\ref{24}$), ($\ref{25}$), we obtain the components of magnetic field in cylindrical coordinates as follows,

\begin{multline}\label{28}
\mathrm{B}_r=\frac{\mu_0  I R (z-z_0)}{4 \pi} \\ 
 \int_0^{2\pi}   \frac{ \cos(\gamma-\alpha)}{[r^2+R^2+ (z-z_0)^2-2rR \cos(\gamma-\alpha)]^{3/2}}\ d\alpha
\end{multline}

\begin{multline}\label{29}
\mathrm{B}_{\gamma}=\frac{\mu_0  I R (z-z_0)}{4 \pi} \\
  \int_0^{2\pi}   \frac{ \sin(\alpha-\gamma)}{[r^2+R^2+ (z-z_0)^2-2rR \cos(\gamma-\alpha)]^{3/2}}\ d\alpha
\end{multline}

\begin{multline}\label{30}
\mathrm{B}_z=\frac{\mu_0  I R }{4 \pi}\\
  \int_0^{2\pi}    \frac{ (R-r\ \cos(\gamma-\alpha)) }{[r^2+R^2+ (z-z_0)^2-2rR\ \cos(\gamma-\alpha)]^{3/2}}\ d\alpha
\end{multline}

It is possible to verify that Eq. (\ref{23}) is satisfied and the Eq. ($\ref{29}$), due to symmetry considerations, is always zero.
We rewrite Eqs. ($\ref{28}$) and ($\ref{30}$) with the definition of a parameter $\xi$ as:

\begin{equation}
\xi (r, R, z)=\frac{2r R}{r^2+R^2+(z-z_0)^2}
\end{equation}

Considering  dimensionless parameters $\eta=\frac{r}{R}$, $M_z=\frac{z}{R}$ and putting $z_0=0$, the above equation becomes:

\begin{equation}
\xi (R, z, \eta)=\frac{2\eta }{1+\eta^2+M_z^2}
\end{equation}

After some manipulations and replacing the above equation we obtain that the radial and longitudinal components of the magnetic field are:

\begin{equation}
\mathrm{B}_r=\frac{\mu_0  I  M_z}{4 \sqrt{2}\pi R  }(\frac{\xi}{\eta})^{3/2}\ I_2(\xi)
\end{equation}

\begin{equation}
\mathrm{B}_z=\frac{\mu_0  I  }{4\sqrt{2} \pi R } (\frac{\xi}{\eta})^{3/2}\ ( I_1(\xi)-  \eta\ I_2(\xi))
\end{equation}
where,

\begin{equation}\label{35}
I_1(\xi)=\int_{0}^{\pi}   \frac{ d\psi}{[1-\xi \ \cos(\psi)]^{3/2}}
\end{equation}

\begin{equation}\label{36}
I_2(\xi)=\int_{0}^{\pi}   \frac{ \cos(\psi)}{[1-\xi \ \cos(\psi)]^{3/2}}\ d\psi
\end{equation}
 and $\psi=\gamma-\alpha$ (we put the reference at $\gamma=0$). In the next section, we first investigate the statement of the problem, then we introduce a method to solve the above integrals.

\section{Statement of the Problem}

As we have mentioned in the previous section, we will first report here the current theorems used in literature for solving the fractional integrals.

Riemann-Liouville fractional integral:

Definition: Let Re $\nu>0$ and f be piecewise continuous and integrable on (0, $\infty$). Then for $z>0$ we define 

\begin{equation}
D_{z-z_0}^{-\alpha} f(z)=\frac{1}{\Gamma(\alpha)} \int_{z_0}^z \frac{f(z)}{(z-z_0)^{-\alpha+1} }dz
\end{equation}
when $z=z_0$, this is called Riemann-Liouville fractional integral of the function $f$ of order $\alpha$.

The Weyl fractional derivatives is used when $z_0$ takes on a singular value from $-\infty$ to $\infty$, and it can be expressed as,

\begin{equation}
D_{z^{-\infty}}^{-\alpha} f(z)=\frac{(-1)^{-\alpha}}{\Gamma(-\alpha)} \int_{z_0}^\infty \frac{f(z)^n}{(z-z_0)^{-\alpha-1} }dz
\end{equation}
and 

\begin{equation}
D_{z^{+\infty}}^{-\alpha} f(z)=\frac{1}{\Gamma(-\alpha)} \int_{-\infty}^{z_0} \frac{f(z)^n}{(z-z_0)^{-\alpha-1} }dz.
\end{equation}

The Caputo fractional derivative \cite{caputo1967linear} is used in order to solve the differential equations without defining the fractional order initial conditions. Caputo's definition is as follows.

\begin{equation}
\small{D_{z-z_0}^{-\alpha} f(z)=\frac{1}{\Gamma(n-\alpha)} \int_{z_0}^z \frac{f(z)^n}{(z-z_0)^{\alpha+1+n} }dz, \ \ n-1<\alpha<n.}
\end{equation}

Notice that all above theorems have a limitation for $\alpha$, and we need a theorem which is defined for all values of $\alpha$. The authors of \cite{lovoie1976fractional} and \cite{osler1970fractional} have studied this kind of integral. The procedure is as follows.

First we recall the Cauchy's integral formula:

\begin{equation}
D^n f(z)=\frac{\Gamma(\alpha +1)}{2\pi i} \oint_C \frac{f(z)}{(z-z_0)^{n+1} }dz
\end{equation}

Let the contour of integration be $\gamma(z_0,z^+)$ in which the branch line for $(z-z_0)^{-\alpha - 1}$ starts from the position z and end at the fixed point $z_0$. The above equation is equivalent to the Riemann-Liouville fractional integral when Re $(\alpha) <0$. We divide the contour $\gamma(z_0,z^+)$  into three contours (see Fig. 2),

\begin{equation}
\gamma(z_0,z^+)=\gamma_1(z_0\rightarrow z)\ U\  \gamma_2 (O)\  U\  \gamma_3(z \rightarrow z_0)
\end{equation}

where,

$\gamma_1(z \rightarrow z_0)$ : line segment from z to $z_0$ ;

$\gamma_2(O)$ : small circle centered at $z_0$;

$\gamma_3(z_0\rightarrow z)$ : line segment from $z_0$ to z.

\begin{figure}[h]
 \begin{center}
\includegraphics[width=0.7\linewidth]{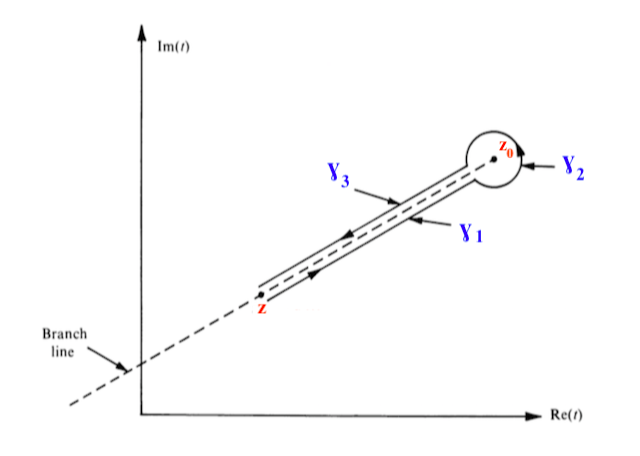}

\caption {contour of integration }
\end{center}

      \end{figure}

Then the Cauchy's integral formula becomes:

\begin{equation}
D^n f(z)=\frac{\Gamma(\alpha +1)}{2\pi i} \int_{\gamma(z_0,z)} \frac{f(z)}{(z-z_0)^{n+1}}dz=I_{\gamma_1}+I_{\gamma_2}+I_{\gamma_3}
\end{equation}

$I_{\gamma_1}, I_{\gamma_2}, I_{\gamma_3}$ denote the integrals over the mentioned contours $\gamma_1, \gamma_2, \gamma_3$. Then, the line in which the branch occurs can be written,

\begin{equation}
\frac{1}{(z-z_0)^{\alpha+1}}= e^{(-\alpha-1)(ln |z-z_0|+i (\theta-\pi))}\ \ \ on\ \ \gamma_1
\end{equation}

\begin{equation}
\frac{1}{(z-z_0)^{\alpha+1}}=0\ \ \ on\ \ \gamma_2
\end{equation}

\begin{equation}
\frac{1}{(z-z_0)^{\alpha+1}}=e^{(-\alpha-1)(ln |z-z_0|+i (\theta+\pi))}\ \ \ on\ \ \gamma_3
\end{equation}

It should be noted that the integral tends to zero on $\gamma_2$ as the contour's radius goes to zero. Substituting the above equations inside of Eq. ($\ref{12}$) we obtain

\begin{equation}
^{\gamma(z,z^+)} D_{z-z_0}^n f(z)=\small{\frac{ (e^{i \pi \alpha}+e^{-i \pi \alpha}) \ \Gamma(\alpha +1)}{2\pi i} \int_{z_0}^z \frac{f(z)}{(z-z_0)^{n+1}}dz}
\end{equation}

or

\begin{equation}
^{\gamma(z,z^+)} D_{z-z_0}^n f(z)=\frac{ sin (\pi \alpha) \ \Gamma(\alpha +1)}{\pi } \int_{z_0}^z \frac{f(z)}{(z-z_0)^{n+1}}dz
\end{equation}

Notice that above equation is valid for all values of $\alpha$ and we can use it for our case as we see in the next section.

\section{Fractional Integral's Solution and the Final Solution}

To solve the integrals $\ref{35}$ and $\ref{36}$ we transform the equations into the complex plane with $z=e^{-i\psi}$ and use the modified Cauchy's integral formula. It should be noted that to use the  Cauchy's integral formula, one should deal with the analytic function of the integral. However, in our case, we have multi-valued operation.

It should be noted that when z goes to $z_0$, it creates a branch line along to the branch point $z_0$. To eliminate the branch line and turn the multi-valued function into an analytic function we divide the contour into three contours in order to use the Cauchy's integral formula [see Appendix]. Such a problem requires an entirely new solution, but the same principle can work also in our case. Finally we obtain

\begin{multline}
I_1(\xi)=\int_{0}^{\pi}   \frac{ d\psi}{[1-\xi \ \cos(\psi)]^{3/2}}\\
=\frac{\pi}{(1+\xi)^{3/2}}\  _2F_1 (\frac{1}{2},\frac{3}{2};1;\frac{2\xi}{1+\xi})
\end{multline}

\begin{multline}
I_2(\xi)=\int_{0}^{\pi}   \frac{ \cos(\psi)}{[1-\xi \ \cos(\psi)]^{3/2}}\ d\psi\\
=\frac{-\pi}{(1+\xi)^{3/2}}\  [\ _2F_1 (\frac{1}{2},\frac{3}{2};1;\frac{2\xi}{1+\xi})- \ _2F_1 (\frac{3}{2},\frac{3}{2};2;\frac{2\xi}{1+\xi})\ ]
\end{multline}

\

Where $\ _2F_1 (\frac{1}{2},\frac{3}{2};1;\frac{2\xi}{1+\xi})$ and $\ _2F_1 (\frac{1}{2},\frac{3}{2};1;\frac{2\xi}{1+\xi})$ are the ordinary hypergeometric functions having a general form of the kind

\begin{equation}
 \ _pF_q(a_1,..,a_p;b_1,..,b_q;z)=\Sigma_{n=0}^\infty \frac{(a_1)_n ... (b_p)}{(b_1)_n...(b_p)_n}\frac{z^n}{n!}
\end{equation}
where $(a_p)_n, (b_q)_n$ are the rising factorial or Pochhammer symbol with

\begin{equation}
(a)_0=1
\end{equation}

and

\begin{equation}
(a)_n=a(a+1)(a+2)...(a+n-1),\ \ \ \ n=1,2,..
\end{equation}

Finally axial and radial magnetic field components are:

\begin{multline}\label{55}
 \mathrm{B}_r=\frac{\mu_0  I \  M_z }{2 R  }(\frac{\xi}{2\eta(1+\xi)})^{3/2}\\
 [\ _2F_1 (\frac{3}{2},\frac{3}{2};2;\frac{2\xi}{1+\xi})\ - \ _2F_1 (\frac{1}{2},\frac{3}{2};1;\frac{2\xi}{1+\xi})]
\end{multline}

\begin{multline}\label{56}
\mathrm{B}_z=\frac{\mu_0  I  }{2 R } (\frac{\xi}{2\eta (1+\xi)})^{3/2}\\
 [(1+\eta) \ _2F_1 (\frac{1}{2},\frac{3}{2};1;\frac{2\xi}{1+\xi})-\eta \ _2F_1 (\frac{3}{2},\frac{3}{2};2;\frac{2\xi}{1+\xi})\ ]
\end{multline}

A simple derivation of the axial magnetic field component can be obtained to evaluate the field on the axis of the coil:

\begin{equation}
\mathrm{B}_z=\frac{\mu_0  I R^2 }{2 (R^2+z^2)^{3/2}} 
\end{equation}

This is a classic equation which can be found in many physics textbooks and papers \cite{jackson1999classical,bassetti1996analytical,amaldi1950fisica}. For the radial on-axis field component ($\eta$=0), both hypergeometric functions of the radial magnetic field components become 1 and cancel each other giving: 

\begin{equation}
\mathrm{B}_r=0
\end{equation}

The authors of \cite{bassetti1996analytical} described the magnetic potential near the axis of a multipole and they have obtained the following equations for the magnetic field components:

\begin{equation}
 \mathrm{B}_r(r, z)=2\Sigma_{p=0}^\infty p\ G_{02p}(z) r^{2p-1} 
\end{equation}

\begin{equation}
\mathrm{B}_z(r, z)=\Sigma_{p=0}^\infty \ G_{02p+1}(z) r^{2p} 
\end{equation}

where,

\begin{equation}
G_{02p}(z)=\frac{\mu_0 I_c}{R^{2p}} \Sigma_{k=0}^\infty F_{0,2p,2k+1}\ f_{2k+1} (z)
\end{equation}

\begin{equation}
G_{02p+1}(z)=\frac{\mu_0 I_c}{R^{2p}} \Sigma_{k=0}^\infty F_{0,2p,2k+1}\  g_{2k+1} (z)
\end{equation}
in the above equations we have the following definitions,

\begin{equation}
f_h (t)=(\frac{z}{\sqrt{R^2+z^2}})^h, \ \ \ \ (h=0, 1, ... \infty)
\end{equation}

\begin{equation}
g_{2k+1} (t)=\frac{df_{2k+1}}{dt} =\frac{(2k+1) R^2}{(R^2+z^2)^{3/2}} f_{2k}(t).
\end{equation}

\begin{equation}
F_{0,2p,2k+1}=\frac{(-1)^p}{4^p (p!)^2} (\mathrm{M}^p \mathrm{F}_{00})_{2k+1}
\end{equation}
where M stands for the second derivative of functions $f_{2k+1}$ and $\mathrm{F_{00}}$ is reported in \cite{bassetti1996analytical}. Taking the first two terms of the magnetic field components we obtain:

\begin{equation}\label{66}
 \mathrm{B}_r(r, z)=\frac{\mu_0  I }{4} [\frac{3R^2 z}{(R^2+z^2)^{5/2}}] r
\end{equation}

\begin{equation}\label{67}
\mathrm{B}_z(r, z)=\frac{\mu_0  I }{2} \ [ \frac{R^2}{(R^2+z^2)^{3/2}} - \frac{3R^2 \ r^2}{2(R^2+z^2)^{5/2}}]
\end{equation}
which coincide with Eqs. $\ref{55}$ and $\ref{56}$ when we consider the first two terms in the expansion of the hypergeometric series $\ _2F_1 (\frac{1}{2},\frac{3}{2};1;\frac{2\xi}{1+\xi})$ and $\ _2F_1 (\frac{1}{2},\frac{3}{2};1;\frac{2\xi}{1+\xi})$.

\

By assuming $(r^2+z^2)^{1/2} \gg R$ and neglecting the higher order terms of R in the denominator of the Eqs. $\ref{55}$ and $\ref{56}$ and finally taking only few terms in the expansion of the hypergeometric functions, the axial and radial magnetic field components can be written as

\begin{equation}
\mathrm{B}_r(r, z)=\frac{\mu_0 m}{4\pi} \frac{3r z}{(r^2+z^2)^{5/2}}
\end{equation}

\begin{equation}
\mathrm{B}_z(r, z)=\frac{\mu_0  m}{4\pi} (\frac{3z^2}{(r^2+z^2)^{5/2}}-\frac{1}{(r^2+z^2)^{3/2}})
\end{equation}
where  $m$ stands for the magnetic moment of the coil, $ m=  I S$, with $S$ its cross section. These are well known expressions of the magnetic field of a small coil (magnetic dipole) at distance much greater than the radius of the loop \cite{mencuccini1988fisica} .

\section{Comparison Between CST Particle Studio Simulation and Analytical Results}

We remember that, since our starting point was the current density on a 2D surface as given by Eq. (\ref{9}), the field components expressed by Eqs. (\ref{55}) and (\ref{56}) are strictly valid for a coil with negligible thickness. In order to check the validity of our results also for a solenoid of finite thickness, in this section we compare the analytical results with those obtained with a 3D electromagnetic code, CST Particle Studio  \cite{CstStudioSuite}.

\begin{figure}[h]
 \begin{center}
\includegraphics[width=1.1\linewidth]{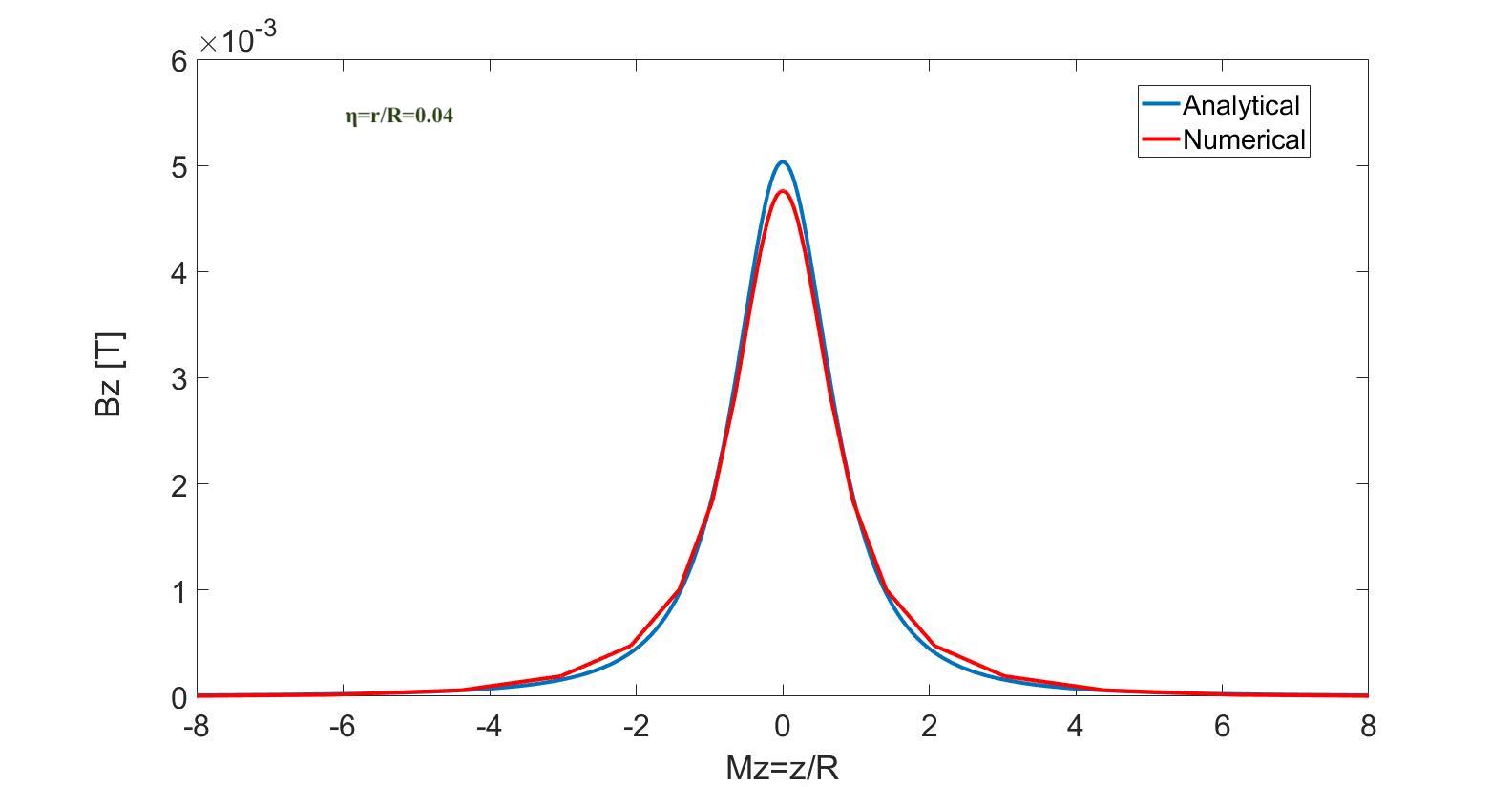}
\includegraphics[width=1.1\linewidth]{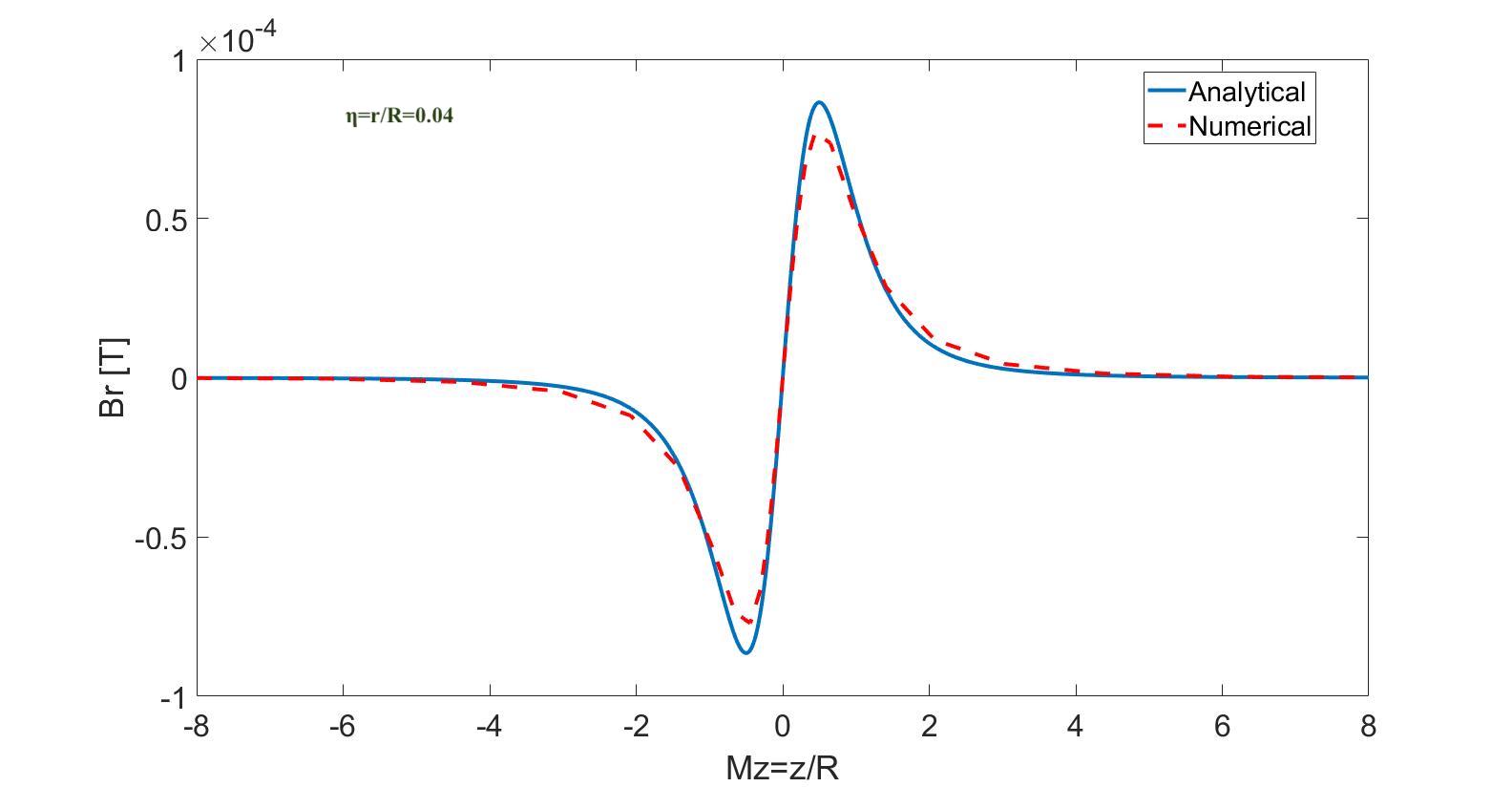}

\caption {Comparison between analytical methods and numerical results of off-axis ($\eta=r/R=0.04$, $r=1$ mm) axial and radial magnetic field components of a solenoid with $R=25$ mm and with a cylindrical cross section of  t= 3 mm. The numerical results have performed by CST Particle Studio. }
\end{center}

      \end{figure}

We have performed comparisons for different thickness  of the solenoid as a function of $M_z=z/R$. In Fig. 3 we have compared analytical method and numerical results for off-axis ($\eta=r/R=0.04$, $r=1$ mm) axial and radial magnetic field components of a solenoid with $R=25$ mm and with a cylindrical cross section with a thickness of $t=3$ mm. The maximum error between analytical and numerical results is about 4$\%$ for the axial magnetic field component and 8$\%$ for the radial magnetic field component. 

In Fig. 4 we have performed the comparisons for an higher off-axis case with $\eta=r/R=0.5$ and $r=12.5$ mm. The normalized axial and radial magnetic fields are shown for two solenoids  with the thicknesses of t=1 mm and 3 mm. The solenoid's radius for both cases was $R=25$ mm.

\begin{figure}[h]
 \begin{center}
\includegraphics[width=1.1\linewidth]{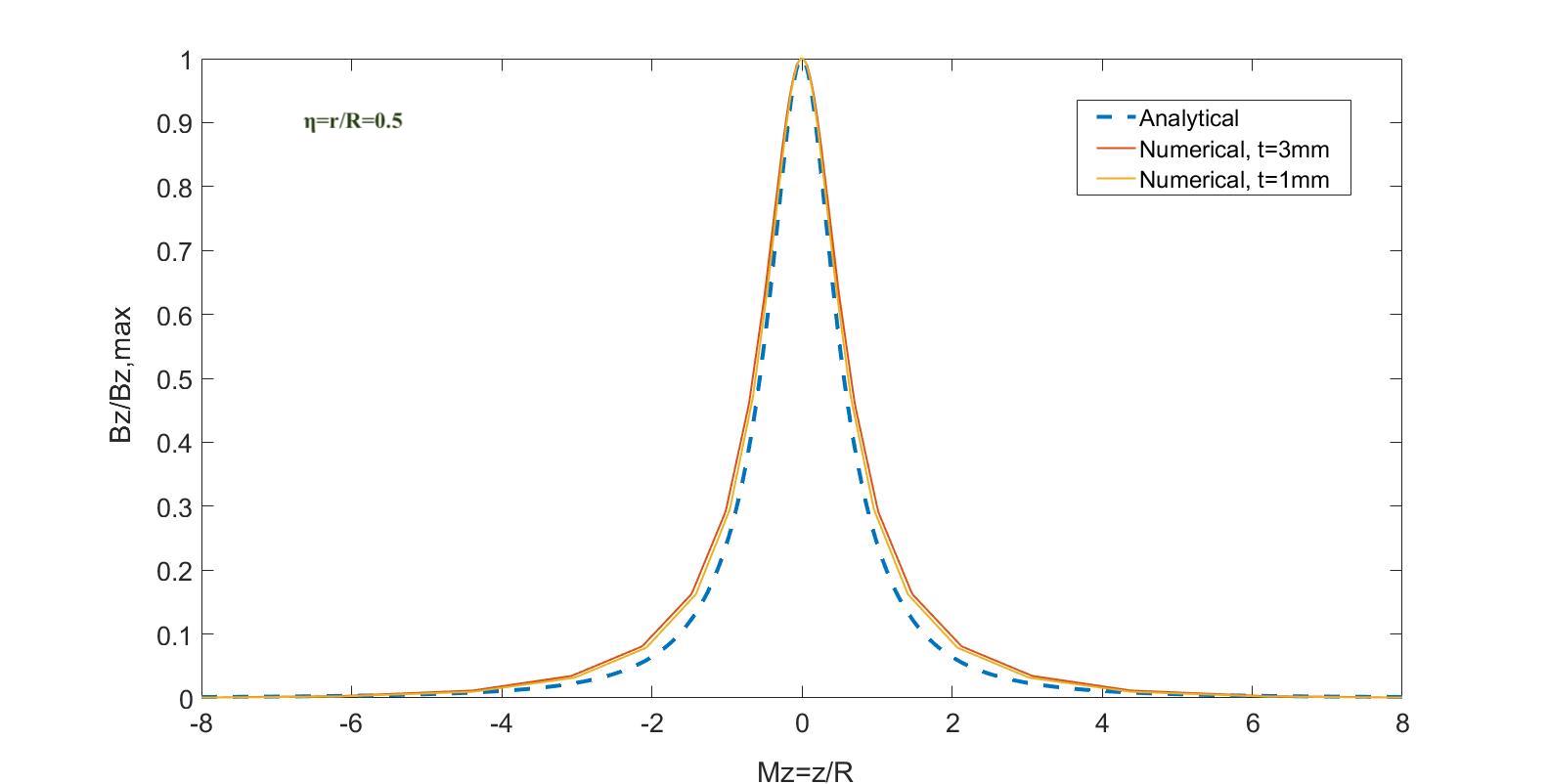}
\includegraphics[width=1.1\linewidth]{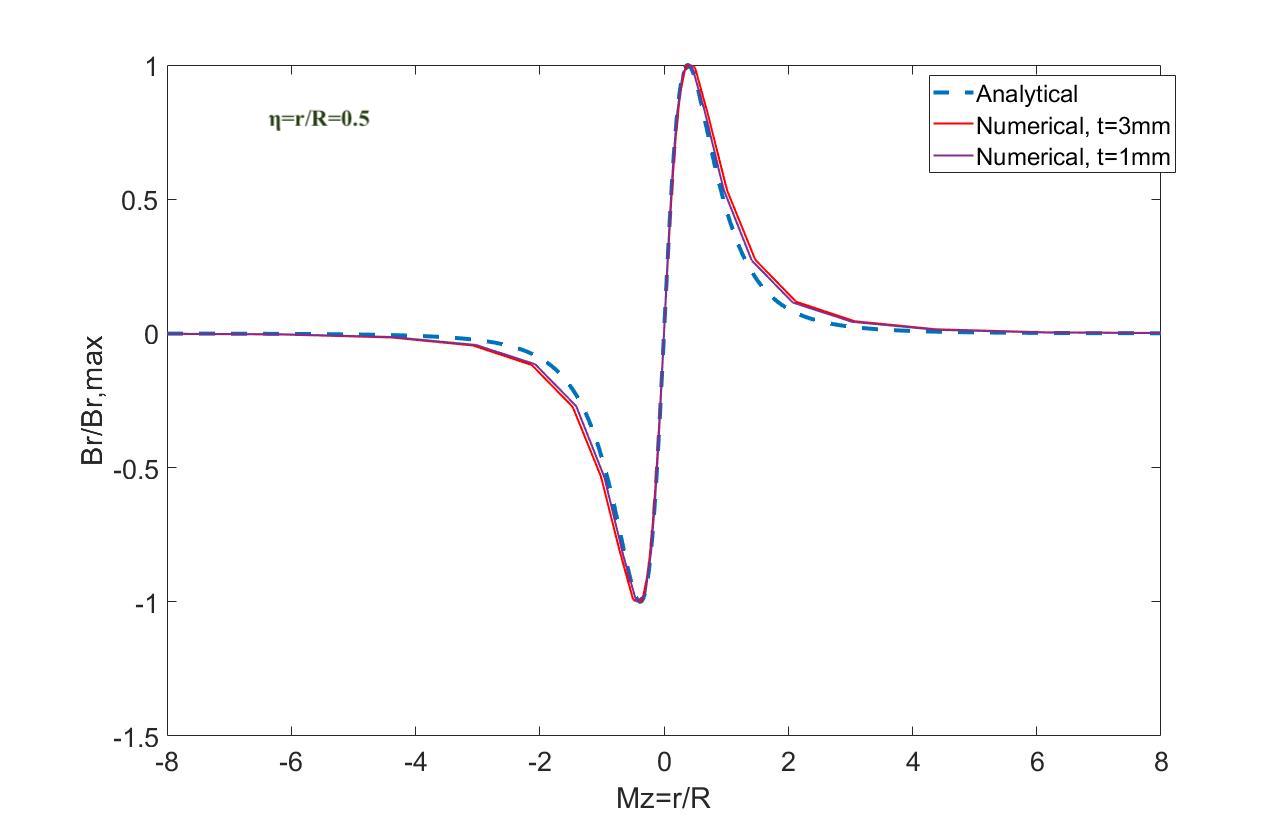}

\caption {Comparison between analytical methods and numerical results of off-axis ($\eta=r/R=0.5$, $r=12.5$ mm) normalized axial and radial magnetic field components for two solenoids with the thicknesses of t=1 mm and 3 mm. The solenoid's radius for both cases was $R=25$ mm.  The numerical results have performed by CST Particle Studio. }
\end{center}

      \end{figure}

      In Figure 5 we show the comparison of the normalized axial magnetic field component of the solenoid for the off axis case of $\eta=r/R=0.5$ and $r=12.5$ mm with three different thicknesses of the solenoid: $t=1$ mm, 3mm and 10mm. As we expect, by decreasing the solenoid's thickness, the error decreases compared with analytical result. However, for all the shown cases, the error is of the order of few percent.

\begin{figure}[h]
 \begin{center}
\includegraphics[width=1.1\linewidth]{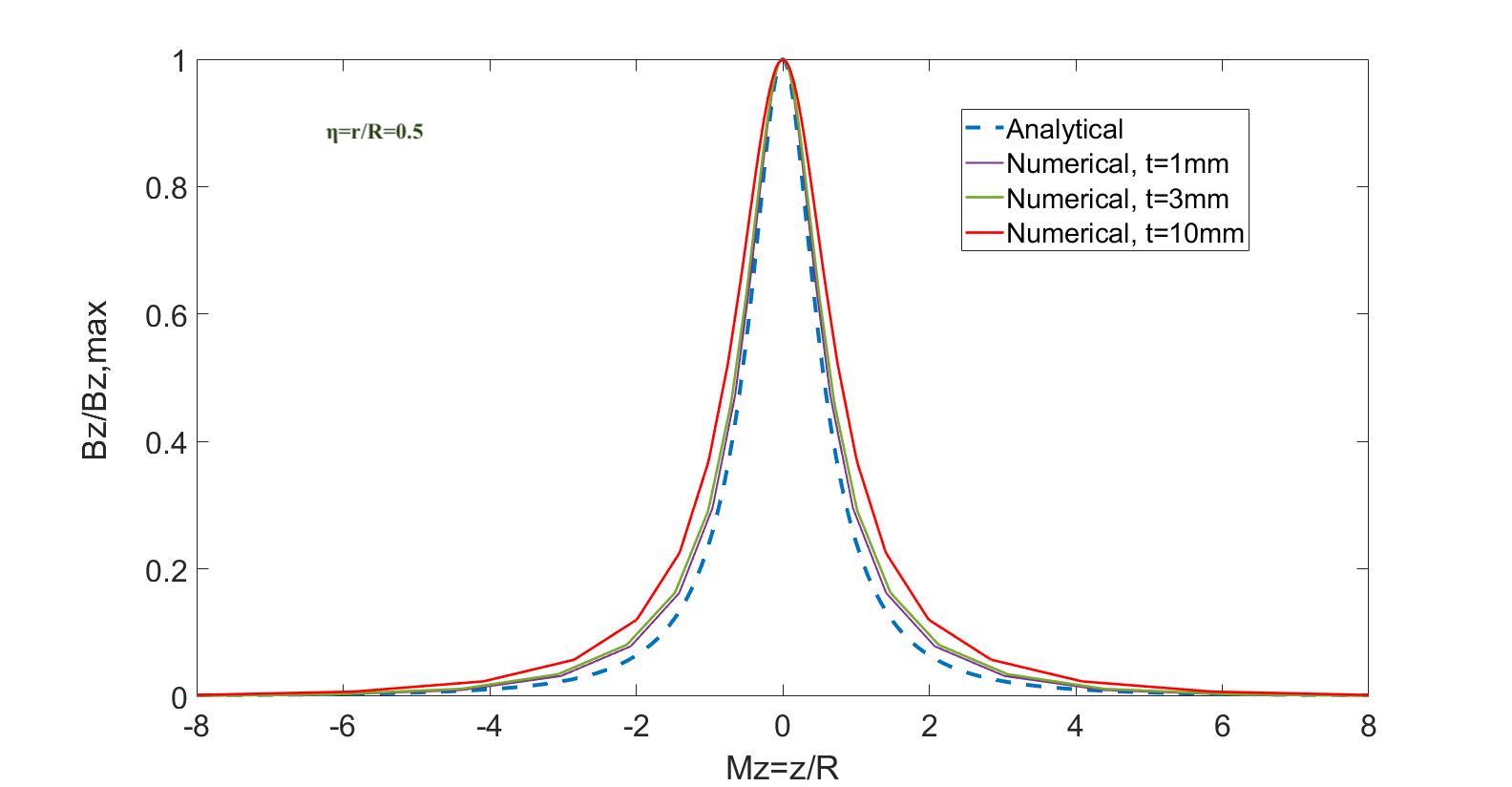}

\caption {Comparison between analytical methods and numerical results of off-axis ($\eta=r/R=0.5$, $r=12.5$ mm) normalized axial magnetic field ($B_z$) for different solenoids with $R=25$ mm and with a cylindrical cross section of $(\omega) t=1$ mm, 3mm and 10mm. The numerical results have performed by CST Particle Studio. }
\end{center}

      \end{figure}

\section{Conclusion}
In this paper, we reported the derivation of the on and off-axes magnetic field components of a solenoid with negligible thickness by means of the hypergeometric functions $_2F_1(a,b;c;z)$. The results have been compared with some known analytical expressions in some simple cases and with the numerical code CST Particle Studio. We obtained a very good agreement. The method described in this paper can be used to obtain the magnetic field of correctors in a LINAC or for the magnetic focusing system of a klystron.

\appendix[Solution of the Fractional Integral]

We begin with Cauchy's integral formula:

\begin{equation}
D^n f(z)=\frac{\Gamma(\alpha +1)}{2\pi i} \oint_C \frac{f(z)}{(z-z_0)^{n+1} }dz
\end{equation}

Let the contour of integration be $\gamma(z_0,z^+)$. The branch line for $(z-z_0)^{-\alpha - 1}$ starts from the position z and ends at the fixed point $z_0$. The above equation is equivalent to the Riemann-Liouville fractional integral when Re $(\alpha) <0$. We divide the contour $\gamma(z_0,z^+)$  into three contours (see Fig. 6),

\begin{equation}
\gamma(z_0,z^+)=\gamma_1(z_0\rightarrow z)\ U\  \gamma_2 (O)\  U\  \gamma_3(z \rightarrow z_0)
\end{equation}

where,

$\gamma_1(z \rightarrow z_0)$ : line segment from z to $z_0$;

$\gamma_2(O)$ : small circle centered at $z_0$;

$\gamma_3(z_0\rightarrow z)$ : line segment from $z_0$ to z.

\begin{figure}[h]
 \begin{center}
\includegraphics[width=0.7\linewidth]{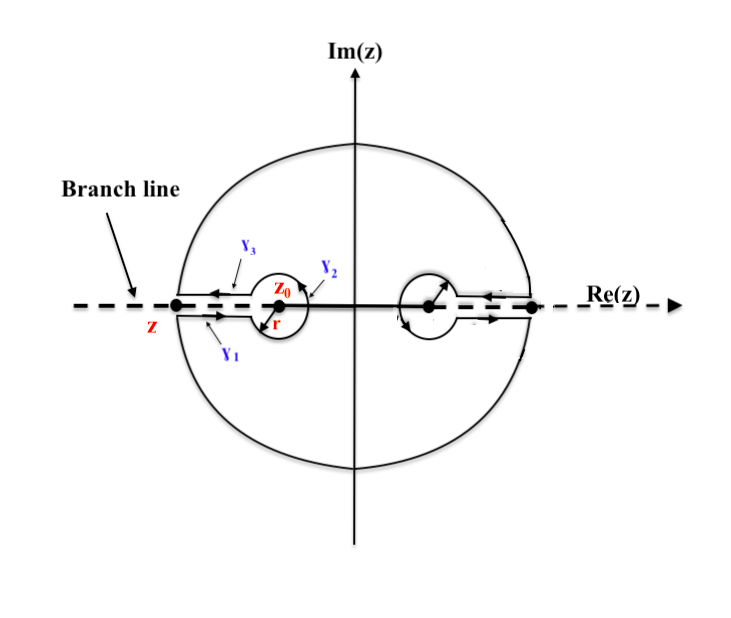}

\caption {contour of integration }
\end{center}

      \end{figure}

Then the Cauchy's integral formula becomes:

\begin{equation}
D^n f(z)=\frac{\Gamma(\alpha +1)}{2\pi i} \int_{\gamma(z_0,z)} \frac{f(z)}{(z-z_0)^{n+1}}dz=I_{\gamma_1}+I_{\gamma_2}+I_{\gamma_3}
\end{equation}
$I_{\gamma_1}, I_{\gamma_2}, I_{\gamma_3}$ denote the integrals over the mentioned contours $\gamma_1, \gamma_2, \gamma_3$. Then, the line in which the branch occurs can be written as

\begin{equation}
\frac{1}{(z-z_0)^{\alpha+1}}= e^{(-\alpha-1)(ln |z-z_0|+i (\theta-\pi))}\ \ \ on\ \ \gamma_1
\end{equation}

\begin{equation}
\frac{1}{(z-z_0)^{\alpha+1}}=0\ \ \ on\ \ \gamma_2
\end{equation}

\begin{equation}
\frac{1}{(z-z_0)^{\alpha+1}}=e^{(-\alpha-1)(ln |z-z_0|+i (\theta+\pi))}\ \ \ on\ \ \gamma_3
\end{equation}

It should be noted that the integral tends to zero on $\gamma_2$ as the contour's radius goes to zero. Substituting the above equations inside of Eq. ($\ref{12}$) we obtain

\begin{equation}
^{\gamma(z,z^+)} D_{z-z_0}^n f(z))=\small{\frac{ (e^{i \pi \alpha}+e^{-i \pi \alpha})  \Gamma(\alpha +1)}{2\pi i} \int_{z_0}^z \frac{f(z)}{(z-z_0)^{n+1}}dz}
\end{equation}

or

\begin{equation}
^{\gamma(z,z^+)} D_{z-z_0}^n f(z)=\frac{ sin (\pi \alpha) \ \Gamma(\alpha +1)}{\pi } \int_{z_0}^z \frac{f(z)}{(z-z_0)^{n+1}}dz.
\end{equation}

Notice that above equation is valid for all values of $\alpha$. Now we have an equation to be used for calculating our integral. Returning to our fractional integral

\begin{equation}
 \int_{0}^{2\pi}   \frac{ 1}{[1-\xi \ \cos(\psi)]^{3/2}}\ d\psi
\end{equation}
by writing the variable $\cos(\psi)$ in the complex plan as $\cos\psi=\frac{z+z^{-1}}{2}$, and replacing into the above equation, after some manipulations we obtain

\begin{equation}
 \int_{0}^{2\pi}   \frac{ 1}{[1-\xi \ \cos(\psi)]^{3/2}}\ d\psi=\small{\frac{2^{3/2}\pi}{\Gamma (3/2)} lim_{z\rightarrow z_0} D_{z}^{1/2}{(z-z_0)^{-3/2}  f(z)}}
\end{equation}

where $z_{01}=\frac{1+\sqrt{1-\xi^2}}{\xi}$ and $z_{02}=\frac{1-\sqrt{1-\xi^2}}{\xi}$ are the branch points of the integral in which the residues should be computed.

\begin{multline}
 \int_{0}^{2\pi}   \frac{ 1}{[1-\xi \ \cos(\psi)]^{3/2}}\ d\psi=2\pi i\  \Sigma_{k=1}^n Res_f(z_k)\\
=2\pi \   \Sigma Res \frac{f(z)}{z}\\
=2\pi \   \Sigma Res \frac{1}{(1-\xi \frac{z+z^{-1}}{2})^{3/2}}\frac{1}{z}\\
=2^{5/2}\pi \   \Sigma Res \frac{z^{1/2}}{\xi^{3/2}(-z^2+2(z/\xi)-1 )^{3/2}}.
\end{multline}

As there is a symmetry in the integral, it does not need to branch cut both the branch lines. For this reason we will take the interval [0,$\pi$] where one of the branch line is located,

\begin{multline}\label{81}
 \int_{0}^{\pi}   \frac{ 1}{[1-\xi \ \cos(\theta)]^{3/2}}\ d\psi=2\pi i\  \Sigma_{k=1}^n Res_f(z_k)\\
=2^{5/2}\frac{\pi}{\Gamma (3/2)} [lim_{z\rightarrow \frac{1-\sqrt{1-\xi^2}}{\xi}} D_{z}^{1/2}(z-z_1)^{3/2} f(z)]\\
=2^{5/2}\frac{\pi}{\Gamma (3/2)} [D_{z-z_1}^{1/2}\frac{z^{1/2}}{(z-(\frac{1-\sqrt{1-\xi^2}}{\xi}))^{3/2}}]
\end{multline}
where $D_{z}^{1/2}$ is the fractional derivative of the order $1/2$. Applying $D_{z}^{1/2}$ to the function we obtain,

\begin{multline}\label{82}
D_{z}^{1/2}\ (\frac{z^{1/2}}{(z-z_0)^{3/2}})\\
=\frac{\Gamma(3/2)}{(z-z_0)^{3/2}}+\frac{z\ \Gamma(5/2)}{2(z-z_0)^{5/2}}-\frac{z^2\ \Gamma(7/2)}{16(z-z_0)^{7/2}}+...
\end{multline}

Finally substituting the Eq. ($\ref{82}$) inside the $($\ref{81}$)$ we obtain:

\begin{multline}
 \int_{0}^{\pi}   \frac{ 1}{[1-\xi \ \cos(\theta)]^{3/2}}\ d\theta= 2^{5/2}\frac{\pi}{\Gamma (3/2)}\\
  [\frac{\Gamma(3/2)}{(z-z_0)^{3/2}}+\frac{z\ \Gamma(5/2)}{2(z-z_0)^{5/2}}-\frac{z^2\ \Gamma(7/2)}{16(z-z_0)^{7/2}}+...]
  \end{multline}
  
  \begin{multline}
=\frac{\pi}{\Gamma (3/2)} [\frac{\Gamma(3/2)}{(1-\xi^2)^{3/4}}+\frac{(1-\sqrt{1-\xi^2})\ \Gamma(5/2)}{4(1-\xi^2)^{5/4}}\\
-\frac{(1-\sqrt{1-\xi^2})^2\ \Gamma(7/2)}{64(1-\xi^2)^{7/4}}+...]
  \end{multline}

Where the integral's solution is a hypergeometric function and it can be written in a compact form as

\begin{equation}
\int_{0}^{\pi}   \frac{ d\psi}{[1-\xi \ cos(\psi)]^{3/2}}=\frac{\pi}{(1+\xi)^{3/2}}\  _2F_1 (\frac{1}{2},\frac{3}{2};1;\frac{2\xi}{1+\xi})
\end{equation}

\bibliographystyle{IEEEtran}
\bibliography{Solenoid_versione6}

%









\end{document}